\documentclass[aps,prl,reprint,superscriptaddress,twocolumn,showpacs]{revtex4-1}

\usepackage{graphicx, amsmath} \usepackage[colorlinks=true,allcolors=blue]{hyperref}

\newcommand{\CRO}{Ca$_3$Ru$_2$O$_7$}

\begin{document}



\title{Existence of electron and hole pockets and partial gap opening in the correlated semimetal Ca$_3$Ru$_2$O$_7$}



\author{Hui Xing} 
\affiliation{Key Laboratory of Artificial Structures and Quantum Control (Ministry of Education) and Shanghai Center for Complex Physics, School of Physics and Astronomy, Shanghai Jiao Tong University, Shanghai 200240, China} 

\author{Libin Wen} 
\affiliation{Key Laboratory of Artificial Structures and Quantum Control (Ministry of Education) and Shanghai Center for Complex Physics, School of Physics and Astronomy, Shanghai Jiao Tong University, Shanghai 200240, China} 

\author{Chenyi Shen} 
\affiliation{Department of Physics, Zhejiang University, Hangzhou 310027, China}

\author{Jiaming He} 
\affiliation{Key Laboratory of Artificial Structures and Quantum Control (Ministry of Education) and Shanghai Center for Complex Physics, School of Physics and Astronomy, Shanghai Jiao Tong University, Shanghai 200240, China} 

\author{Xinxin Cai} \affiliation{Department of Physics and Materials Research
Institute, Pennsylvania State University, University Park, PA 16802, U.S.A.} 

\author{Jin Peng}
\affiliation{Department of Physics, Tulane University, New Orleans, LA 70118, U.S.A.}

\author{Shun Wang} \affiliation{Key Laboratory of Artificial Structures and Quantum Control (Ministry of Education) and Shanghai Center for Complex Physics, School of Physics and Astronomy, Shanghai Jiao Tong University, Shanghai 200240, China}

\author{Mingliang Tian} \affiliation{High Magnetic Field Laboratory, Chinese Academy of Sciences,
Hefei 230031, China} \affiliation{Collaborative Innovation Center of Advanced Microstructures,
Nanjing 210093, China.} 

\author{Zhu-An Xu}
\affiliation{Department of Physics, Zhejiang University, Hangzhou 310027, China}
\affiliation{Collaborative Innovation Center of Advanced Microstructures, Nanjing 210093, China.}

\author{Wei Ku} \affiliation{Key Laboratory of Artificial Structures and Quantum Control (Ministry of Education) and Shanghai Center for Complex Physics, School of Physics and Astronomy, Shanghai Jiao Tong University, Shanghai 200240, China} 

\author{Zhi-Qiang Mao}
\email{zmao@tulane.edu}
\affiliation{Department of Physics, Tulane University, New Orleans, LA 70118, U.S.A.} 

\author{Ying Liu}
\email{yxl15@psu.edu}
\affiliation{Key Laboratory of Artificial Structures and Quantum Control (Ministry of Education) and Shanghai Center for Complex Physics, School of Physics and Astronomy, Shanghai Jiao Tong University, Shanghai 200240, China}
\affiliation{Department of Physics and Materials Research Institute, Pennsylvania State University,
University Park, PA 16802, U.S.A.} \affiliation{Collaborative Innovation Center of Advanced
Microstructures, Nanjing 210093, China.} 

\date{\today}

\begin{abstract}
The electronic band structure of correlated \CRO\ featuring an antiferromagnetic as well as a structural transition has been determined theoretically at high temperatures, which has led to the understanding of the remarkable properties of \CRO\ such as the bulk spin valve effects. However, its band structure and Fermi surface (FS) below the structural transition have not been resolved even though a FS consisting of electron pockets was found experimentally. Here we report magneto electrical transport and thermoelectric measurements with the electric current and temperature gradient directed along $a$ and $b$ axes of an untwined single crystal of \CRO\, respectively. The thermopower obtained along the two crystal axes were found to show opposite signs at low temperatures, demonstrating the presence of both electron and hole pockets on the FS. In addition, how the FS evolves across $T^* = 30$ K at which a distinct transition from coherent to incoherent behavior occurs was also inferred - the Hall and Nernst coefficient results suggest a temperature and momentum dependent partial gap opening in \CRO\ below the structural transition, with a possible Lifshitz transition occurring at $T^*$. The experimental demonstration of a correlated semimetal ground state in \CRO\ calls for further theoretical studies of this remarkable material.    
 

\end{abstract} 

\maketitle

Layered ruthenates in the Ruddlesden-Popper series family (Sr, Ca)$_{n+1}$Ru$_n$O$_{3n+1}$ \cite{Ruddlesden1958} have attracted great attention in condensed matter and materials physics community because they were found to show a wide range of exciting  phenomena, including spin-triplet superconductivity in Sr$_2$RuO$_4$ \cite{Mackenzie2003, Kidwingira2006, Liu2015}, band-dependent Mott metal-insulator transition \cite{Mizokawa2001, Nakamura2013} and orbital ordering \cite{Zegkinoglou2005} in Ca$_2$RuO$_4$, metamagnetism, and correlated effects in Sr$_3$Ru$_2$O$_7$ \cite{Perry2001, Borzi2007, Bruin2013}, making them a canonical complex transition metal oxide system for the search of new physical phenomena. The evolution of physics in the Ruddlesden-Popper family (Sr, Ca)$_{n+1}$Ru$_n$O$_{3n+1}$ through the reduction of cation radius, marked by the change of the system from the quantum magnet Sr$_3$Ru$_2$O$_7$ to antiferromagnetic metal Ca$_3$Ru$_2$O$_7$ \cite{Qu2009}, as well as the increase in the number of perovskite RuO$_2$ layers that leads to the transition from a band-dependent Mott insulator Ca$_2$RuO$_4$ to the metallic Ca$_3$Ru$_2$O$_7$ with a $k$-dependent gap \cite{Puchkov1998, Snow2002}, is particularly interesting.

Ca$_3$Ru$_2$O$_7$ was found to show a paramagnetic (PM) metal to antiferromagnetic (AFM) metal transition at $T_N$ = 56 K \cite{Cao1997}. For 48 K $< T < T_N$, the AFM state is characterized by ferromagnetic bilayers stacked antiferromagnetically along the $c$-axis, with the magnetic moments aligned along the $a$-axis. As the temperature is lowered below $T_s = 48$ K,  the system exhibits a first-order phase transition, characterized by the switching of magnetic moments from the $a$- to the $b$-axis \cite{Bao2008} and multiple other changes. Although the orthorhombic crystal symmetry (space group of Bb2$_1$m) remains unchanged through the first-order transition at $T_s$, the structural transition is marked by a clear change in the lattice parameters: the $c$-axis lattice constant is shortened, while those of the $a$- and $b$-axis are enlarged. Such lattice parameter changes are accompanied by the enhanced rotation and tilting of RuO$_6$ octahedra below $T_s$ \cite{Yoshida2005}, as illustrated in Fig. 1(a,b). Interestingly, the first-order phase transition at $T_s$ is also accompanied by a sharp increase in the in-plane resistivity $\rho_{ab}$ \cite{Yoshida2004}, followed by a negative d$\rho_{ab}$/d$T$, identified previously as a metal-insulator transition. Additionally, a dramatic bulk spin-valve phenomenon was discovered \cite{Cao2008} and understood based on the unusual itinerary magnetic state \cite{Singh2006}. 
	
\begin{figure*}[t] \centering \includegraphics[width=1.6\columnwidth]{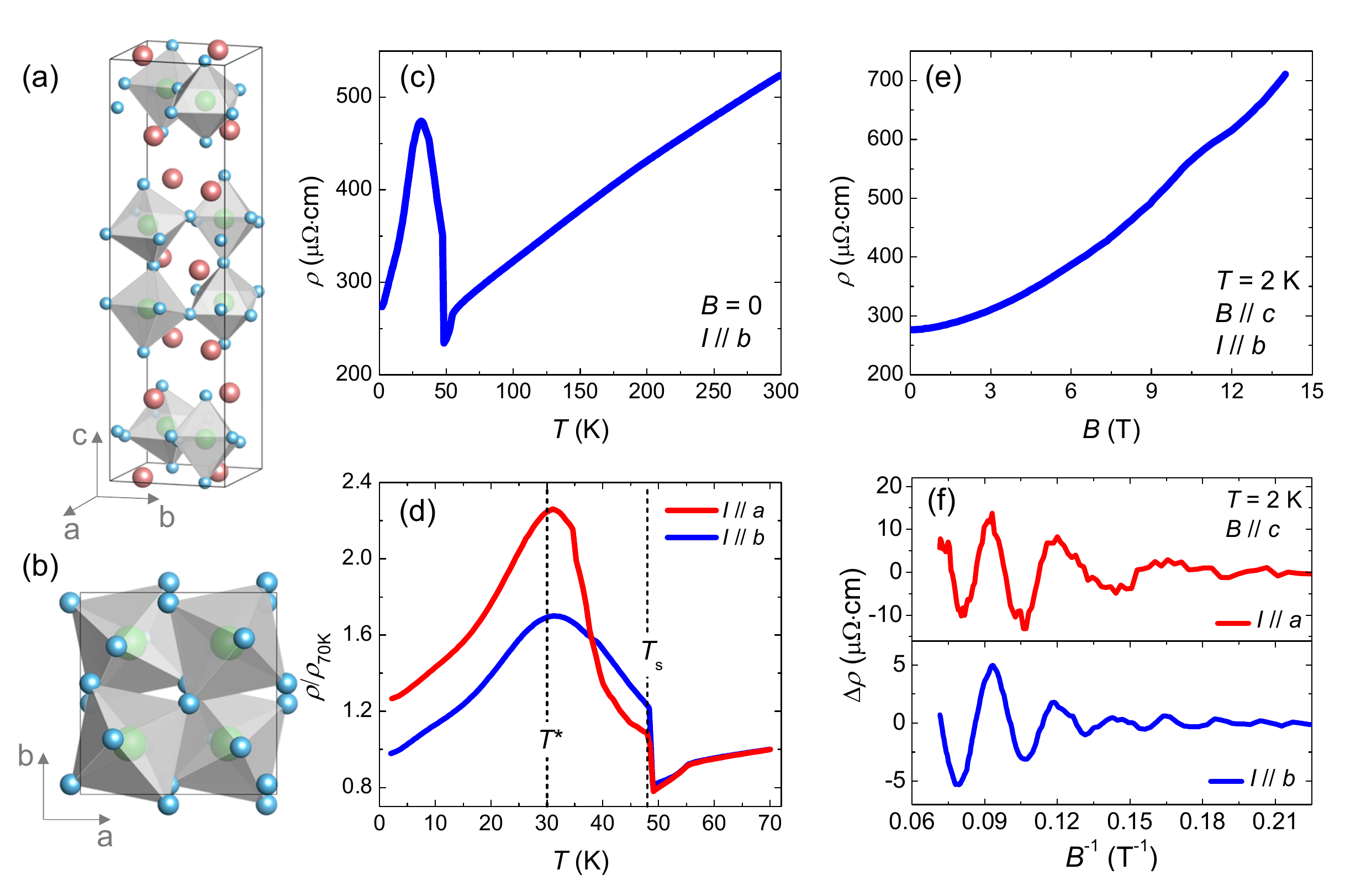} 
\caption{(a) Crystal structure of \CRO. Red, green and blue balls stand for Ca, Ru and Oxygen,
respectively. (b) The top view of the unit cell, with Ca cations neglected. (c) Zero-field
resistivity of \CRO\ along the $b$ axis as a function of temperature. (d) Temperature dependence of
normalized resistivity along $a$ and $b$ axes. (e) Field dependence of $b$-axis direction
resistivity at 2 K for magnetic field along the $c$ axis. (f) Shubnikov-de Haas oscillations at 2 K
along $a$ and $b$ axes with magnetic field along the $c$ axis obtained from $\rho(H)$ by subtracting
smooth background.} \end{figure*}

Electronic band structures of correlated metals such as \CRO\, which serves as a useful starting point to understand its physical properties, can be calculated if the correlated effects are dealt with properly and verified experimentally. Even though the band structure calculations of \CRO\ were attempted \cite{Liu2011}, no results consistent with experimental results have been reported. On the other hand, a tight-binding argument suggests the presence of hole pockets in addition to electron ones \cite{Kikugawa2010}. Experimentally, ARPES also revealed the presence of small electron pockets at low temperatures, which will account for the observed in-plane metallic behavior. Previous quantum oscillations measurements have yielded partly inconsistent results in the presence of multiple frequencies \cite{Cao2003, Baumberger2006, Kikugawa2010}. In addition, $\rho_{ab}$($T$) was found to become metallic below $T^* = 30$ K, raising questions on whether the ``insulating" state for  $T^* < T < T_s$ is actually metallic possessing a FS and the origin of change from the incoherent to coherent behavior at around $T^*$. It also raises an interesting question on the nature of this first-order phase transition at $T_s$ to begin with. In this regard, the opening of a density wave at $T_s$ was suggested based on optical spectroscopic studies \cite{Lee2007}. No direct evidence for the presence of a density wave has been found in \CRO\, however. In this regard, even a momentum dependent gap was indeed observed in ARPES measurements below $T_s$ \cite{Baumberger2006}, the FS cannot be determined by either the quantum oscillations or ARPES measurements at such high temperatures. All this calls for alternative methods to determine the FS. Here, using orientation dependent magneto electrical and thermoelectric transport measurements, we find the first experimental evidence for the presence of both electron and hole pockets and partial gap opening.

\begin{figure}[t] \centering \includegraphics[width=1.0\columnwidth]{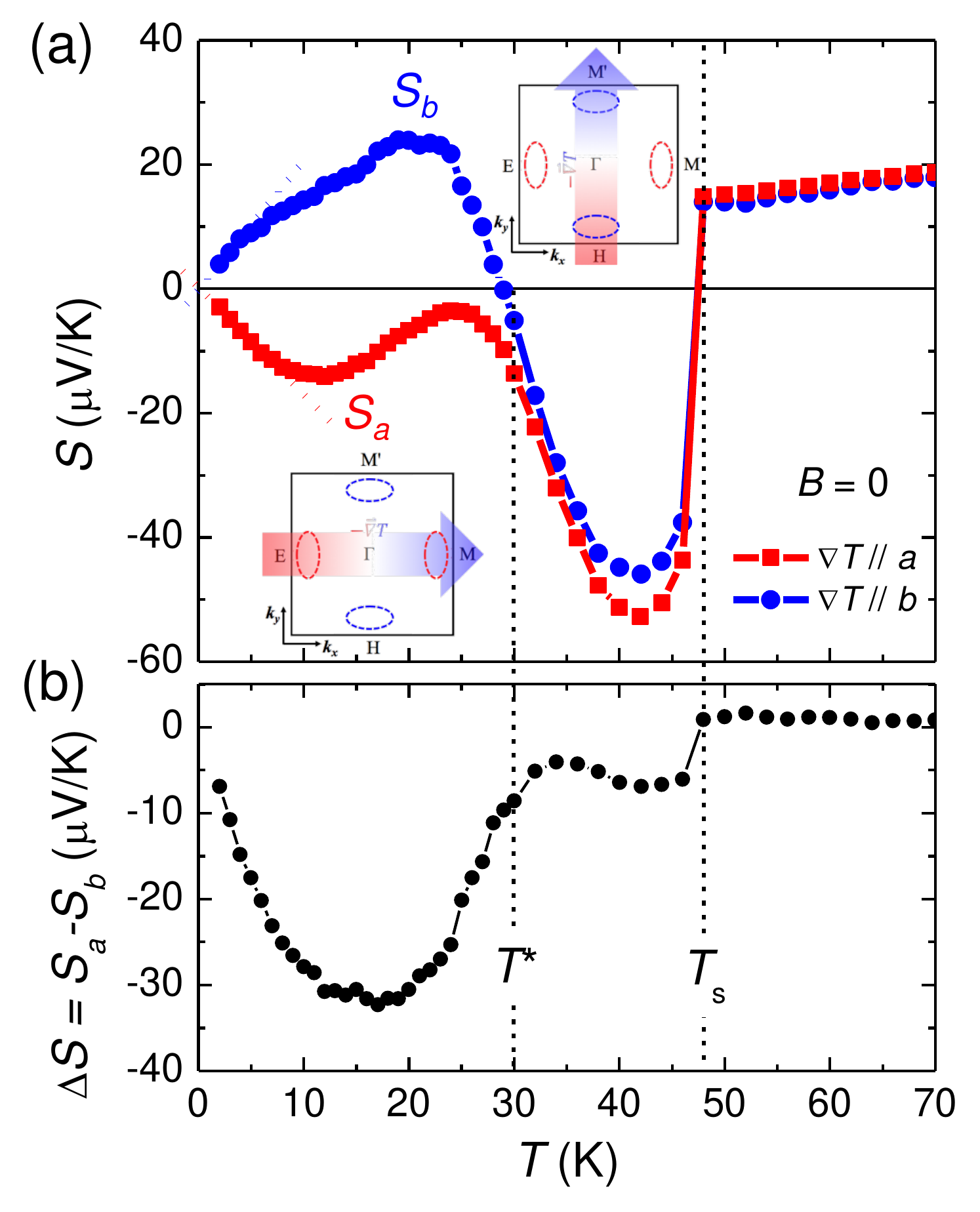} 
\caption{(a) Zero-field thermopower of \CRO\ along $a$ and $b$ axes as a function of temperature.
Inset schematics show a low-temperature Fermi surface schematic adopted from
\cite{Baumberger2006}, the direction of the thermal gradient in the $k$ space; (b) The temperature dependence of 
thermopower anisotropy, defined as $\Delta S = S_a-S_b$.} \end{figure}

Single crystals of \CRO\ were grown by floating zone technique. To probe physics related to the in-plane anisotropy, 
it is critical to use clean twin-free crystals. For this, we performed systematic screening procedure using X-ray diffraction, Laue diffraction and SQUID magnetometry to identify clean twin-free crystals. 
Selected crystals were cut along the $a$ and $b$ axis, respectively, with a rectangular shape. Resistivity, Hall and 
thermoelectric measurements were performed in a Quantum Design PPMS system with a 14 Tesla magnet.
A steady-state technique was used in thermoelectric measurements. The direction of $-\nabla T$ 
relative to directions in the first Brillouin zone is shown schematically by the arrows in the insets of Fig. 2. 
This allows perturbation of part of Fermi surfaces with Fermi velocity parallel to $-\nabla T$. For systems with anisotropic electronic states, this method can be a sensitive probe in complementary to well established probes such as Shubnikov-de Hass oscillation and ARPES measurements. 

\begin{figure}[t] \centering \includegraphics[width=1.0\columnwidth]{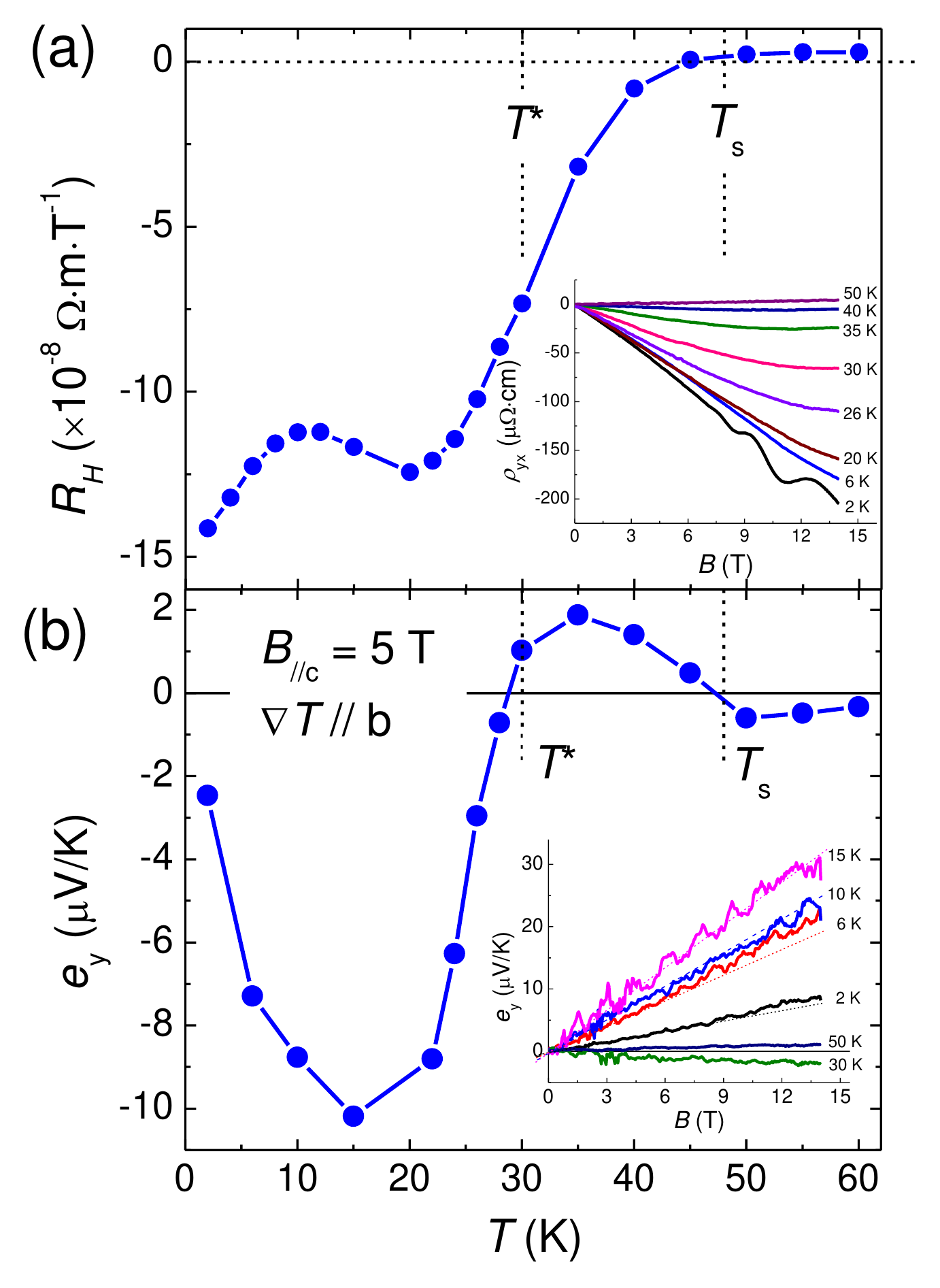} \caption{(a) Temperature
dependence of Hall coefficient in \CRO. Inset shows the field dependence of Hall resistivity at
various temperatures; (b) Temperature dependence of Nernst signal $e_y$ in \CRO. Inset shows the field
dependence of the Nernst signal. Thermal gradient is along the $b$ axis. 
} \end{figure}

At high temperatures, \CRO\ was found to feature metallic behavior, as shown in zero-field resistivity
data obtained in a sample prepared by a $b$-axis crystal in Fig. 1(c). A change in slope appeared 
at $T_N$, corresponding to the onset of the AFM transition. Upon further cooling, a sharp jump in resistivity 
was found, along with a negative slope in $\rho(T)$, as seen previously in the in-plane resistivity measurements 
with an unspecified in-plane current direction \cite{Yoshida2005}. At $T^*$, resistivity values obtained along both 
$a$- and $b$-axis were found to show an incoherent-to-coherent crossover (Fig. 1(d)), well above the 
temperature at which a similar transition was found in the $c$-axis resistivity (at $T = 8$ K). 
Magnetoresistance (MR) at low temperatures was found to show Shubnikov-de Haas oscillations (SdHOs),
as seen in the $\rho(H)$ curve in Fig. 1(e). The oscillatory part in $\rho(H)$, obtained by
subtracting a smooth background, is plotted in Fig. 1(f).  The periodicity in the
$\Delta\rho(B^{-1})$ for both $a$ and $b$-axis resistivity gives a frequency in SdHOs of $\sim 41$ T,
suggesting rather tiny FS pockets, $\sim 1\%$ BZ, a value consistent with those found previously \cite{Cao2003, Kikugawa2010, Puls2014}. 

\begin{figure}[t] \centering \includegraphics[width=1.0\columnwidth]{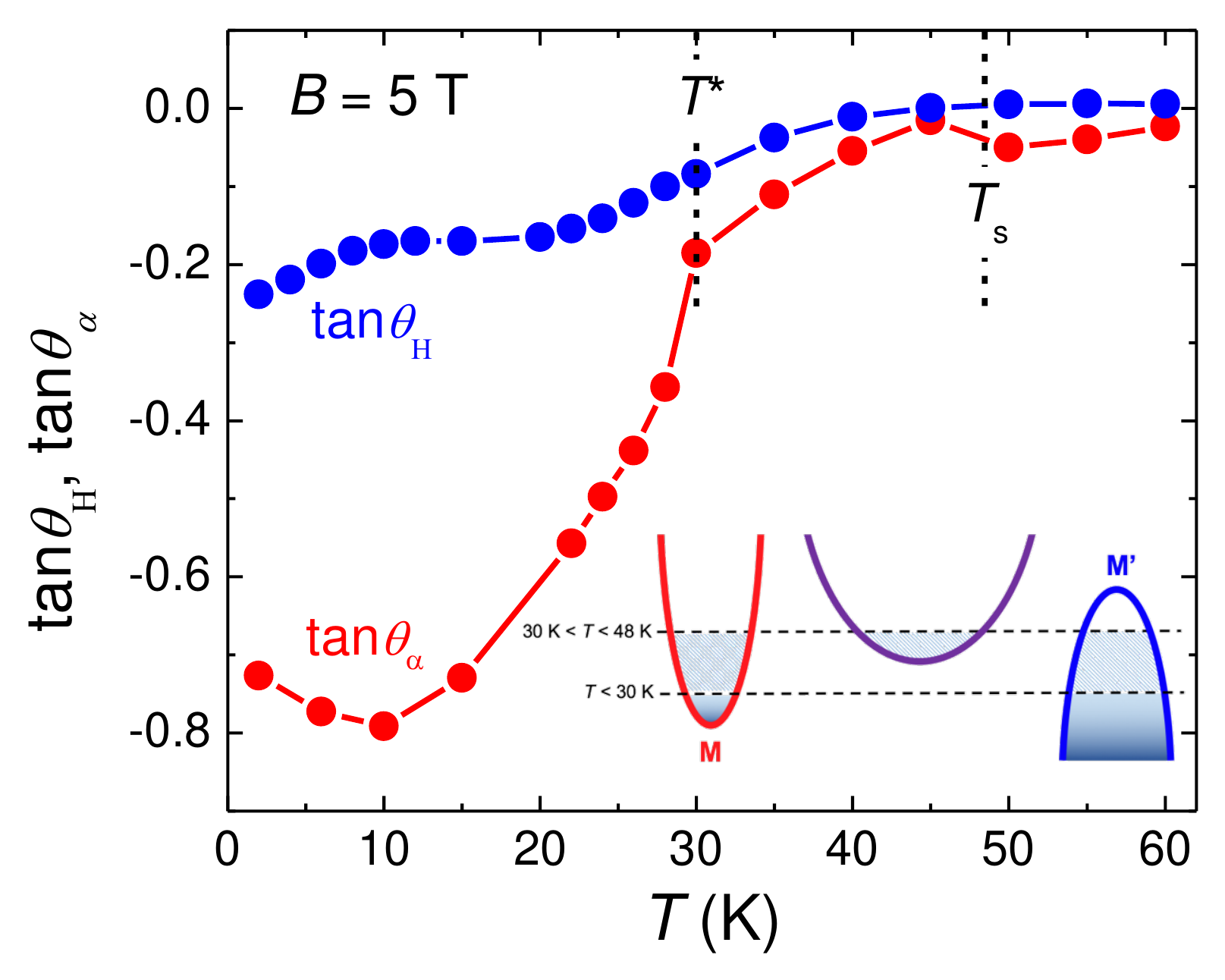} 
\caption{The temperature dependence of
thermal Hall tan$\theta_{\alpha}$ and Hall angle tan$\theta_{H}$ at 5 T. Inset: a schematic showing a possible
Lifshitz transition driven by the shift of chemical potential. Below the structural transition,  
the band structure of \CRO\ consists of an electron and hole band at $M$ and $M'$ point, 
and another electron-like band. The chemical potential decreases with
lowering temperature, and misses the large electron-like band when cooling below $T^*$.} \end{figure}

Thermopower data measured with the temperature gradient along $a$ and $b$ axes, denoted as $S_a$ and $S_b$, 
respectively, are shown in Fig. 2. It is seen that both $S_a$ and $S_b$ are positive and nearly identical at high temperatures. 
With the decreasing temperature, the thermopower was found to decrease, but no signature was found at the magnetic transition 
around $T_N$. At $T_s$, a sharp drop was found in both $S_a$ and $S_b$, with the difference between the two 
becoming significant.  At around $T^*$ = 30 K, $S_b$ changed sign to positive while $S_a$ remains negative. The magnitudes 
of both $S_a$ and $S_b$ were seen to decrease in the low-temperature limit, as required for an entropy current. 

Consider now the implication of the thermopower data. Neglecting correlation effects, thermopower of electrons 
can be expressed in terms of conductivity \cite{Blatt1976, Barnard1972}, %
\begin{equation}\label{eqn:S}
\begin{aligned} 
     S=\frac{\alpha_{xx}}{\sigma_{xx}}&=-\frac{\pi^{2}k_{B}^{2}T}{3|e|} \frac{\partial \ln\sigma}{\partial E}\bigg\vert_{E_{F}} \\
      &=-\frac{\pi^{2}k_{B}^{2}T}{3|e|}\left[\frac{1}{A}\frac{\partial A}{\partial E}+\frac{1}{l}\frac{\partial l}{\partial E}\right]\bigg\vert_{E_{F}}
\end{aligned}
\end{equation}
where $\sigma$ denotes the conductivity, $\alpha$ the Peltier conductivity, $A$ the FS
area, and $l$ the carrier mean free path. One can see that thermopower is therefore a measure of the variation in
conductivity with respect to chemical potential. In general, the second term in the square brackets in eq.\ref{eqn:S} is much smaller than the first term, therefore the sign of thermopower is related directly to the carrier type of the dominating band. 
	
As discussed above, earlier works on quantum oscillation, ARPES and band structure calculation has set up clear boundary condition: tiny FS consisting possible features around M and M' points (electron- and hole-like pockets) \cite{Baumberger2006, Kikugawa2010}. In our measurement, the thermal gradient $-\nabla T$ was directed towards 
the $\Gamma\text{-}M$ and $\Gamma\text{-}M'$ direction in the $k$ space for $S_a$ and $S_b$, respectively. 
The negative $S_a$ and positive $S_b$ seen at low temperatures therefore indicate the existence of a dominating electron and hole bands in their respective directions: the first evidence supporting the presence 
of both electron and hole pockets in \CRO. In a one-band nearly-free-electron approximation \cite{Behnia2009},
$S=\frac{\pi^2}{2}\frac{k_{B}}{e}\frac{T}{T_{F}}$, the slope of $S(T)$ at low-temperature provides an estimate of the Fermi temperature, leading to $T_{F}^+ = 350$ K for the hole pocket and for the electron pocket, $T_{F}^- = 425$ K. The rather low Fermi temperatures are expected for a low carrier density system. 
On the other hand, it is also important to note that in the presence of subtle FS structures, for instance, a van Hove singularity near the FS \cite{McIntosh1996}, or complex FS curvatures \cite{Ong1991}, the sign of thermopower cannot be linked to the type of carrier directly. However, these special cases do not seem to occur in our system according to earlier ARPES and band structure calculation as discussed above. 

Values of $\Delta S = S_a - S_b$, a quantitative measure of the thermopower anisotropy, plotted in Fig. 2(b), suggests strongly the presence of two regimes below $T_s$ = 48 K. The sharp change also indicates that additional change occurs at around 30 K, which cannot be accounted for by the gapped bands at 48 K as described in an earlier thermopower measured with an arbitrary in-plane direction \cite{Iwata2007}. To understand the nature of the electronic state in these two regimes, as well as that of the incoherent-coherent 
crossover found at $T^*$ = 30 K, we investigated further the Hall and Nernst effect in \CRO. The Hall resistivity $\rho_{H}$ in the
inset of Fig. 3(a) is seen to depend on the field linearly at low fields with nonlinearity seen 
at high fields, which is attributed to multiband effects. The Hall 
coefficient $R_{H}$ shown in Fig. 3(a) reveals a sign change at the first-order phase transition at
$T_s$, and nonmonotonic temperature dependence at lower temperatures. 
The sharp increase in the magnitude of $R_H$ suggests a rapid growth in $l_e/l_h$, and thus a significant reduction in 
scattering in the electron-like bands \cite{Ong1991}. It is worth to note that similar behavior has been found in several two-dimensional CDW systems featuring saddle points on the FS \cite{Rice1975}. Whether this applies to \CRO\ is yet to be verified. 

The Nernst signal $e_y = E_y/\nabla T$ measures the transverse electric field $E_y$ generated by a longitudinal temperature gradient $-\nabla T$ in the presence of magnetic field. Here $e_y$ was measured with $-\nabla T$ along $b$ axis and was found to depend on the field linearly at low fields. 
Nonlinearity is seen at high fields (inset in Fig. 3(b)). In addition, the temperature dependence of $e_y$ shown in Fig. 3(b)
features a sign change at $T_s$ and slightly below $T^*$. A drastic enhancement in its magnitude was found below $T^*$, reaching a value 
as large as 10 $\mu$V/K at 15 K. A large $e_y$ can arise from several possible sources \cite{Wang2001, Behnia2009}. In the two-band picture%
\begin{equation}\label{eqn:ey}
e_y =
S(\frac{\alpha^{+}_{xy}+\alpha^{-}_{xy}}{\alpha^{+}_{xx}+\alpha^{-}_{xx}}-\frac{\sigma^{+}_{xy}+\sigma^{-}_{xy}}{\sigma^{+}_{xx}+\sigma^{-}_{xx}}).
\end{equation}
where $\alpha$ is the Peltier conductivity tensor with the sign of carriers denoted by the superscript, ``+" or ``-" for the holes and the electrons, 
respectively. We note that from the Hall coefficient $R_H(T)$, the system is compensated at around $T_s$,
i.e., $\sigma^+_{xy} = -\sigma^-_{xy}$, leading to a vanished second term in eq. \ref{eqn:ey}.
Therefore, a sizable change in $e_y$ around $T_s$ is expected. $e_y$ was also found to change sign
at $T_s$, indicating that the first term in $e_y$ is comparable to the second term
around this temperature. For the same reason, $e_y$ remains
relatively small. Below $T^*$, $R_H$ becomes increasingly negative, 
while $e_y$ is negative and large in magnitude. Therefore the sharp decrease in $e_y$ must come from a strong
reduction in the first term, $i.e.$, the off-diagonal Peltier coefficient term, which points to a
change in scattering rate for $T < T^*$. 
This change in the scattering rate is further demonstrated in Fig. 4. Here we compare the temperature dependence of Hall angle 
$\tan\theta_H = \sigma_{xy}/\sigma_{xx}$ and the Peltier angle $\tan\theta_\alpha = \alpha_{xy}/\alpha_{xx}$. The former corresponds 
to the carrier mobility therefore probes the scattering time while the latter is sensitive to the energy dependence of the scattering time \cite{Wang2001}. 
It is seen that $\tan\theta_H$ features an increase in its magnitude below $T_s$ but no anomaly around $T^*$. On the other hand, the Peltier angle 
is seen to show a rapid increase at $T^*$. The large Peltier angle, nearly four times bigger than the Hall angle, suggests a significant change 
in the energy dependence of the conductance for $T < T^*$. 

The above measurements would suggest a momentum dependent gap opening below $T_s$ and the electronic state of Ca$_3$Ru$_2$O$_7$ experiences a significant change at around $T^*$. For $T^* < T < T_s$, limited part of the FS is gapped out, leaving thermopower taken with $-\nabla T$ along both $\Gamma\text{-}M$  and $\Gamma\text{-}M'$ directions dominated by an electron-like band. As a result, both $S_a$ and $S_b$ are negative and the anisotropy $\Delta S$ is small. Below $T^*$, however, most of the electron-like band on the FS is gapped out, but the electron and hole pockets near M and M' points survive. Furthermore, the temperature dependent gap opening occurs gradually over a large temperature range as the temperature is lowered below $T_s$, which explains the absence of a clear signature at $T^*$ in specific heat data \cite{McCall2003, Yoshida2004}.

It is likely that the density wave formation is responsible for the momentum dependent gap opening, provided that the nesting condition for the density wave varies with the temperature. This will result in a temperature dependent gap opening. The existing ARPES data appears to support this scenario. In this regard, a small jump in the $k\text{-}$dependent gap was found around $T^*$ in earlier ARPES data (see Fig. 4(c) in Ref. \cite{Baumberger2006}), which not only supports the temperature-dependent nesting condition picture but also suggests a Lifshitz transition driven by a shift in the chemical potential as the temperature is lowered, as depicted in the inset of Fig. 4. In this picture, the change in the nesting condition is abrupt at $T^*$.
Lifshitz transition describes the change of Fermi surface topology without breaking any symmetry of the system. The continuous change of an order parameters, as found in traditional phase transitions, no longer exist. Instead, the topological invariants dictate the transition.
Incidentally, an appreciable change in chemical potential was indeed found to exist in several semimetals \cite{Wu2015, Brouet2013}. 
It is known that a Lifshitz transition will affect the material property significantly due to the reconstructed FS, especially in materials with magnetic or charge instabilities. For example, the nesting condition was found to change significantly at the Lifshitz transition in pnictide superconductors \cite{Dhaka2013}. A similar situation may be encountered in \CRO. 

In summary, we provide the first experimental evidence for the existence of both electron and hole pockets in \CRO\ at low temperatures through the measurement of anisotropic thermopower. Furthermore, from the measurement of Hall and Nernst coefficient, we found evidence for a partial gap opening in an extended temperature range below $T_s$. These findings help in resolving the standing issue on the low-temperature Fermi surface configuration and provide new insight for further understanding of the intricate behavior of \CRO\ at around $T^*$. 

\begin{acknowledgments} The authors have benefited from discussion with Anthony Leggett, Yan Chen,
Hong Sun and Dong Qian. The work done at SJTU was supported by MOST (Grant No. 2015CB921104), NSFC (Grant Nos. 91421304 and 11474198) and the Fundamental Research Funds for the Central Universities, at Penn State by NSF (Grant No. EFMA1433378), at ZJU was supported by NSFC under Nos. U1332209 and 11774305, at CAS by NSFC Grant No.U1432251 and the CAS/SAFEA international partnership program for creative research teams of China, at Tulane supported by the U.S. Department of Energy under EPSCoR Grant No. DE-SC0012432 with additional support from the Louisiana Board of Regents. \end{acknowledgments}


\bibliographystyle{apsrev4-1}

\end{document}